\let\myo\o

\documentclass{article}

\usepackage{graphicx}
\usepackage{color}

\usepackage{amssymb,amsfonts,amsmath}

\begin{document}

\title{Machine learning for predictive condensed-phase simulation}

\author{Albert P. Bart\'ok\footnote{Engineering Laboratory, University of Cambridge, Trumpington Street, Cambridge, CB2 1PZ, United Kingdom}, Michael J. Gillan\footnote{London Centre for Nanotechnology, University College London, Gordon St., London WC1H 0AH, United Kingdom; Department of Physics and Astronomy, University College London, Gower St., London WC1E 6BT, United Kingdom; Thomas Young Centre, University College London, Gordon St., London WC1H 0AH, United Kingom}, Frederick R. Manby\footnote{Centre for Computational Chemistry, School of Chemistry, University of Bristol, Bristol BS8 1TS, United Kingdom}
\and G\'abor Cs\'anyi$^*$}

\maketitle


\begin{abstract}

We show how machine learning techniques based on Bayesian inference
can be used to reach new levels of realism in the computer
simulation of molecular materials, focusing here on water.
We train our machine-learning algorithm using accurate, correlated quantum chemistry,
and predict energies and forces
in molecular aggregates ranging from clusters to solid
and liquid phases. The widely used electronic-structure
methods based on density-functional theory (DFT) give poor accuracy
for molecular materials like water, and we show how our
techniques can be used to generate systematically improvable
corrections to DFT. The resulting corrected DFT scheme gives remarkably accurate
predictions for the relative energies of small water clusters and of different ice
structures, and greatly improves the description of the structure and dynamics of liquid water.
\end{abstract}

\section{Introduction}
\label{sec:intro}

The computer simulation of materials has become an indispensable tool across a wide range of
disciplines, including chemistry, metallurgy, the earth sciences, surface 
science and biology.
Simulation techniques range all the way from simple empirical force fields to the
electronic structure methods based on density functional theory (DFT) and correlated
quantum chemistry~\cite{MatMod}. Electronic structure methods are capable of much greater accuracy
and generality than force fields, but their computational demands are
heavier by many orders of magnitude. A crucial challenge for simulation is therefore
to find systematically improvable methods for casting information from accurate
electronic-structure techniques into forms that are more rapidly computable. 
We show here how machine learning techniques~\cite{Mackay} allow this to be done
using correlated quantum chemistry for molecular materials, taking condensed-phase
water as our example.

The fundamental interactions in water and other molecular materials~\cite{stone1996} consist of
exchange-repulsion, electrostatic interaction between molecular charge distributions,
polarization (i.e. the electrostatic distortion of charge distributions), charge transfer
and van der Waals dispersion, together with effects due to molecular flexibility.
Electron correlation plays a role in all these, and is crucial for dispersion~\cite{klimes2012}. The
correlated quantum chemistry methods of MP2 (2nd-order M${\rm\myo}$ller-Plesset) and
particularly CCSD(T) (coupled-cluster with single and double
excitations and perturbative triples)~\cite{helgaker2000} give a very accurate description of these
interactions~\cite{tschumper2002,bates2009}, but their heavy computational demands for extended
systems make their routine use for condensed
matter problematic. DFT techniques are less demanding, and
have been widely used for water~\cite{laasonen1993}, but it has been found that the results with standard
approximations often agree poorly with experiment~\cite{grossman2004} and may depend strongly
on the assumed approximation~\cite{wang2011}. There is
vigorous debate about how to overcome the problems, and we believe that input from
correlated quantum chemistry is essential. In the approach described here, machine
learning~\cite{Mackay} is used to construct representations of the energy differences between correlated quantum
chemistry and DFT, which can then be used to construct efficient corrected DFT schemes for simulation of large, complex systems.

Our machine-learning methods are based on the recently reported ideas of
Gaussian Approximation Potentials (GAP)~\cite{bartok09,bartok10}. For molecular materials, we use these
ideas in the framework of the widely used many-body representation~\cite{xantheas1994,pedulla1996}, in which the
total energy $E_{\rm tot} ( 1, 2, \ldots N )$ of a system of $N$ molecules is separated
into 1-body, 2-body and beyond-2-body parts:
\begin{equation}
E_{\rm tot} ( 1, \ldots N ) = \sum_{i=1}^N E_{\rm 1B} ( i ) +
\sum_{i < j} E_{\rm 2B} ( i, j ) + E_{\rm B2B} ( 1, \ldots N) \; .
\end{equation}
Here, $E_{\rm 1B} ( i )$ is the 1-body energy of molecule $i$ in free space, which depends
on its distortion away from its equilibrium configuration. The energy $E_{\rm 2B} ( i, j )$
is the 2-body interaction energy of the pair of molecules $( i, j )$ in free space, i.e.
the total energy of the pair minus the sum of their 1-body energies. For water,
$E_{\rm 2B} ( i, j )$ is a function of 12 variables specifying the separation of the
molecules, their relative orientation and their internal distortions. The beyond-2-body (B2B)
energy $E_{\rm B2B}$ represents everything not accounted for by 1- and 2-body energies.
Exchange-repulsion, 1st-order electrostatics and dispersion are mainly or entirely 2-body
interactions, while $E_{\rm B2B}$ arises mainly from polarization and perhaps charge transfer,
with contributions from exchange-repulsion and dispersion expected to be smaller~\cite{wang2010}.
The ideas in the present paper build on the pioneering works in the groups of Szalewicz~\cite{bukowski2007} and Bowman~\cite{Bowman}. 

The energy of an isolated H$_2$O monomer as a function of distortion is known from \emph{ab initio} calculations to extremely high precision. We use the parameterization due to Partridge and Schwenke (PS)~\cite{partridge1997}, which can be regarded as essentially exact for our purposes.
In water and many other molecular materials, the 2-body interactions give the largest contribution
to the binding energy, so these must be accurately represented~\cite{hodges1997}. However, in water, the B2B 
interactions also play a crucial role~\cite{bukowski2007}. It has long been known that there is a strong redistribution
of electrons when a water monomer enters the liquid or solid phases, and its dipole
moment increases from $1.8$~D in the gas phase to 
$\sim 2.5$~D or more in condensed phases~\cite{coulson1966}. The B2B
terms in the energy come partly from the large changes of electrostatic interactions due to
this redistribution. In the strategy presented here, the B2B part of the energy is represented by
a chosen form of DFT, because DFT can be expected to include most of the physical effects that
play a role in the B2B energy. We will discuss at the end how such an approximation to B2B 
interactions can be further improved. We note here that we fully include molecular flexibility,
which is not always done even in sophisticated empirical interaction models.
Flexibility is essential, because without it one could not describe the well known lengthening
of O-H bonds and the lowering of O-H vibrational frequency~\cite{jeffrey1997} when the H atom participates
in a hydrogen bond with another monomer, and nor could quantum nuclear 
effects~\cite{habershon2009} be properly treated.

\section{Machine learning with GAP}
\label{sec:machine_learning}

Consider a system whose configuration is specified by points ${\bf R}$ in
a many-dimensional configuration space. We are given the values $f ( {\bf R}_n )$ of its
energy at a finite set of configurations $\{{\bf R}_n\}$. We now ask:
what is the most likely value of $f ( {\bf R} )$ at a configuration ${\bf R}$ not in the given
set $\{ {\bf R}_n \}$? The rules of Bayesian inference~\cite{Mackay} are used to compute this most likely
value, assuming that the function $f$ has certain smoothness properties. Here smoothness simply means
that the probability of finding very different values $f ( {\bf R} )$ and $f ( {\bf R}^\prime )$  
decreases rapidly to zero as ${\bf R}$ and ${\bf R}^\prime$ approach each other. The
framework of GAP, based on Gaussian processes\cite{GP}, uses a precise formulation of smoothness in terms of a covariance function $C ( {\bf R} , {\bf R}^\prime )$ having the form~\cite{GP}:
\begin{equation}
C ( {\bf R} , {\bf R}^\prime ) = \theta \exp\left({-\sum_i [(R_i-R_i^\prime)/(2\sigma_i)]^2}\right) ,
\end{equation}
where the sum in the exponent is over the dimensions of the configuration space, $\theta$ is the typical scale of $f$ and  $\sigma$ is the typical length scale on which $f ( {\bf R} )$ varies. The theory yields the following formula~\cite{GP} for the most likely estimate of $f ( {\bf R} )$ given the data and the assumption of smoothness
(often called the maximum \emph{a posteriori} estimator):
\begin{equation}
f ( {\bf R} ) = \sum_n^{\rm data} C ( {\bf R} , {\bf R}_n ) \alpha_n \; ,
\end{equation}
where the coefficients $\alpha_n$ are given by inversion of the linear equations:
\begin{equation}
f ( {\bf R}_m ) = \sum_n^{\rm data} [ C ( {\bf R}_m , {\bf R}_n ) + \varepsilon \delta_{mn}]\alpha_n \;,
\end{equation}
where $\delta_{mn}$ is the Kronecker delta; the diagonal shift of magnitude $\varepsilon$ is included to
regularise the linear algebra. 

When applying GAP to represent 1- and 2-body energetics in water,
there are different ways of choosing the space of points ${\bf R}$ representing configurations,
but here it is advantageous to build in the fact that the energy function $f ( {\bf R} )$
is left unchanged by rotations and translations of the whole system, and by interchange of
identical atoms. For the water monomer, the two OH distances and the angle between them provide
a convenient coordinate system. For the water dimer, we ensure rotation and translation symmetry by working with the space of
the 15 interatomic distances, ${\bf R} = \{ | {\bf r}_i - {\bf r}_j | \}$, where ${\bf r}_i$ are
the atomic positions. To ensure interchange symmetry, we symmetrize the covariance
function over permutations of identical atoms:
\begin{equation}
\tilde C(\mathbf{R},\mathbf{R}^\prime) = \frac1{|S|}\sum_{\pi\in S}C(\pi(\mathbf{R}),\mathbf{R}^\prime),
\end{equation}
where $S$ is the permutation group of the water dimer, whose order $ |S| $  is 8. A more detailed account 
of our GAP formalism is given in~\cite{bartok09} and \cite{bartok10}. The computational cost of evaluating the GAP model is linear in the size of the 
database $\{{\bf R}_n\}$, and for the present case takes about 10 ms on a single processor. 

Recently, Gaussian processes were used to model the atomisation energies of small molecules\cite{rupp2012}, and a similar technique (based on neural networks) was used to fit the total energy of the water dimer in the DFT approximation~\cite{BehlerWater}. 
Rather than using GAP to represent the whole dimer energy, we use it to represent
the difference between basis-set converged CCSD(T) and DFT, in other words the
DFT error. In practice, we do this in two stages. In the first stage, we compute the difference
between MP2 and DFT at about ten thousand dimer configurations, using
the AVTZ basis. In using GAP to represent this difference, we also use the gradients of 
the potential energy surface (see SI). In the second stage, we construct a GAP model using 
around one thousand energies (without forces) to represent 
the small difference between MP2/AVTZ and CCSD(T), the latter converged with respect to the basis set to about 1 meV.

In the calculations to be presented, we use GAP to correct the popular Becke-Lee-Yang-Parr (BLYP)
approximation of DFT, our choice being guided by the fact that BLYP is accurate for the B2B energy
of small water clusters (see SI). We show in Fig.~1 the 2-body errors of BLYP together with the errors of GAP-corrected BLYP for a thermal sample of dimer configurations (which were not used in the construction the GAP model) drawn from a molecular dynamics simulation of liquid water.
Uncorrected BLYP is too repulsive for the water dimer, with unacceptably
large errors of up to $50$~meV at the separations of interest. However, with GAP corrections,
the errors are dramatically reduced to $\sim 1$~meV. GAP thus provides a way of virtually eliminating
all errors in a chosen DFT approximation apart from those associated with B2B energy. Also shown
in Fig.~1 are the errors of the approximation obtained by the popular procedure of adding
the dispersion correction due to Grimme \emph{et al.}~\cite{grimme2010} to BLYP. We note that this
approximation is better than uncorrected BLYP, but is much less good than GAP-corrected BLYP.


\section{Results}
\label{sec:results}

To illustrate the power of our GAP-corrected DFT, we start with a simple test
on the ten stationary points of the water dimer~\cite{smith1990}. These form a canonical
set of configurations,
which have been exhaustively studied and whose energies
are extremely accurately known~\cite{tschumper2002}. The global minimum structure is bound by
a single hydrogen bond, but some of the less stable structures have up to
four weaker hydrogen bonds. (For pictures of the structures, see e.g. Ref.~\cite{tschumper2002}.)
We stress that none of these stationary points is
included in our training set, so that the energies computed with BLYP~$+$~GAP
are genuine predictions. We compare in Table~1
the relative energies of the 10 configurations from BLYP~$+$~GAP
with the almost exact results from Ref.~\cite{tschumper2002} and the predictions of the DFT
functional BLYP; the Table also includes the very accurate predictions of diffusion
Monte Carlo (DMC)~\cite{gillan2012}. As has been reported before~\cite{anderson2006}, the DFT approximation
shows quite large errors of around $30$~meV
in some cases,
while the errors of DMC are much smaller, being almost all less than $3$~meV. 
Our BLYP~$+$~GAP predictions are very accurate, and 
indeed they compete in accuracy with DMC, at enormously reduced computational cost.

As a second test, we examine the predictions of $\text{BLYP} + \text{GAP}$ for the energies of
different isomers of the water hexamer. This system has been studied extensively for many 
years~\cite{pedulla1996,gregory1997,bates2009},
for a very important reason. The most stable structures of small water clusters from the trimer to the pentamer
have a ring-like form in which each monomer is hydrogen bonded to two neighbours~\cite{gregory1997}. However, from
the hexamer onwards, rings are less stable than compact structures in which some monomers are hydrogen
bonded to three or four neighbours~\cite{gregory1997,kim1999,bates2009}. The energy balance for the hexamer is rather delicate, but
high-precision CCSD(T) calculations leave no doubt that the compact prism and cage structures
have lower energy than the more open book and ring forms~\cite{bates2009}. However, many of the commonly
used DFT approximations, including BLYP and PBE, wrongly predict that the ring or book form is
most stable~\cite{santra2008}. We compare in Figure~\ref{fig:hex} the predictions of $\text{BLYP} + \text{GAP}$ with
CCSD(T) benchmarks and with the predictions of BLYP and DMC. We see that again the GAP-corrected DFT model is highly accurate
and is comparable to DMC.

For any material, crystal energetics provides a crucial test
of modelling techniques. Water has a remarkably rich phase diagram, with no fewer
than fifteen known ice structures~\cite{petrenko1999,salzmann2009}. In 
the common form ice Ih, found at ambient pressure,
each H$_2$O monomer is H-bonded to four nearest neighbours at an O-O distance
of 2.75~\AA, the next-nearest neighbours having the much greater O-O separation
of 4.5~\AA. The pattern of H-bonding in ice Ih is disordered, but the 
closely related ordered form ice XI,
stable below 72~K, has essentially the same local geometry~\cite{petrenko1999}. With increasing pressure,
denser structures become more stable, and we will be concerned here with 
(in order of increasing density) ice IX, II, XV and VIII. The distances to 
non-H-bonded next-nearest neighbours decrease along this series, becoming almost
exactly equal to the first-neighbour distance in ice VIII~\cite{petrenko1999}. There are accurate
experimental values for the zero-pressure energies and volumes of almost all
these structures.


Standard DFT approximations perform poorly for ice~\cite{santra2011}, making the energies increase far
too much from ice Ih to VIII, and giving transition pressures too high by up to a factor of 10.
Our machine-learning techniques allow us to correct any DFT approximation for
1- and 2-body errors, and enable us to discover whether these errors are responsible
for the poor description of ice energetics. We have used $\text{BLYP} + \text{GAP}$ to
calculate the relaxed geometries and the equilibrium energies and volumes of the ice 
structures mentioned above, substituting the periodic Bernal and Fowler structure for the proton-disordered Ih~\cite{bernalfowler1933}. The results are reported in Table~\ref{tab:ice}, where we also
give results obtained with uncorrected BLYP. The energies and volumes relative
to ice Ih are very accurately given by $\text{BLYP} + \text{GAP}$, whereas the relative
energies with uncorrected BLYP suffer the large errors reported earlier for
standard DFT methods~\cite{santra2011}. However, our results also reveal a significant {\em uniform} overbinding
in all the structures due to beyond-2-body errors, implying that 
$\text{BLYP} + \text{GAP}$ overestimates the strength of cooperative H-bonding, leading to an overestimate of the equilibrium density by about 5--10\%.
The systematic overestimation of density by $\text{BLYP}+\text{GAP}$ would be partially compensated by zero-point effects, which increase volumes by between 1--5\%~\cite{galli}.

Early attempts to elucidate the properties of liquid water on the molecular level
using DFT were promising~\cite{laasonen1993}, but it has turned out that the standard methods
give surprisingly poor predictions~\cite{grossman2004}, for reasons that are
presumably linked to the problems found with ice structures. Other molecular liquids
seem to suffer from related difficulties~\cite{mcgrath2011}. Some of the common approximate functionals
give an equilibrium density of water that is too low by as much as $30$\% (PBE and BLYP
underestimate it by $\sim$10\% and 10--20\% respectively)~\cite{wang2011,tuckerman2012}. In DFT simulations
of the liquid at the experimental density, the diffusion coefficient may be too low
by as much as a factor of $10$, and the liquid is significantly overstructured~\cite{wang2011}
as compared with experimental neutron and x-ray diffraction data. Quite apart
from the contradictions with experiment, it is now clear that the DFT simulations may
be afflicted by many technical sources of error, including system size effects,
basis-set incompleteness, incorrect temperature control, and the neglect of
quantum nuclear effects. Because of the controversies, we approach the liquid
with caution, particularly since our machine-learning methods for going beyond DFT
do not yet account for errors in B2B interactions. Nevertheless, we can
address an important and well defined question: With machine-learning used to
ensure the correctness of 1B and 2B parts of the energy, can DFT errors in the B2B energies
still cause contradictions with experiment? 

We performed molecular dynamics simulations on $64$ molecules of liquid heavy water (D$_2$O) with $\text{BLYP} + \text{GAP}$ at temperature $308$~K and density $1.109$~g~cm$^{-3}$.
The thermal average pressure in the simulation was $ -2.6$~kbar. By contrast,
a simulation based on BLYP itself under exactly the same conditions gives the larger
pressure of $7$~kbar, which is associated with the 10--20\% underestimate
of equilibrium density. A simple estimate based on our observed pressure together with
the experimental compressibility indicates that the equilibrium density with $\text{BLYP} + \text{GAP}$
is higher than the correct value by $\sim$~10\%, which is consistent with the uniform overbinding observed
above for the ice structures. 

We compute the self-diffusion coefficient $D$ of molecules in our simulation in the conventional
way~\cite{frenkel2002} from the slope of the time-dependent mean-square displacement (for details, see SI),
finding the value $1.3 \times 10^{-9}$~m$^2$~s$^{-1}$. In a system of only $64$ 
molecules, the value of $D$ is expected to be reduced by size effects by an amount that can be estimated
by standard methods~\cite{duenweg1993}. Our size-corrected value
(see SI) of $1.7 \times 10^{-9}$~m$^2$~s$^{-1}$ should be compared with
the experimental value of $2.4 \times 10^{-9}$~m$^2$~s$^{-1}$~\cite{hardy2001}.
This contrasts with the values reported for BLYP itself, which are up to an order magnitude too small~\cite{wang2011,Jonchiere2011,tuckerman2007}. Once again, 2-body effects appear to be the main culprit in making BLYP unrealistic. 

The well known DFT errors of overstructuring in liquid water are most clearly seen in the
oxygen-oxygen radial distribution function $g_{\rm OO} ( r )$. A comparison of
the experimental and computed RDFs (using $\text{BLYP} + \text{GAP}$ and uncorrected BLYP --- see Fig.~\ref{fig:goo}) shows that the GAP correction very significantly improves agreement with experiment. The overstructuring of the liquid is largely corrected:
the first peak in $g_{\rm OO} ( r )$ is lowered by $\sim$~0.25 and the
first trough becomes shallower by $\sim$~0.2. However, our comparison with 
experimental data indicates that the liquid is probably still slightly overstructured even with
$\text{BLYP} + \text{GAP}$. We have chosen to compare here with a  joint refinement of
both neutron and x-ray data~\cite{soper2007}, but one should note the extensive discussion in the literature
about the uncertainties in both kinds of experimental data. We attempt to give a balanced summary of
this discussion in the SI. In addition, since our simulation treats the nuclei as classical
particles, it is essential to consider quantum nuclear corrections. Our discussion of the
experimental and theoretical evidence about these (see SI) indicates that they may
lower the first peak height of $g_{\rm OO} ( r )$ by  $\sim$~0.1. Our assessment
that water simulated with $\text{BLYP} + \text{GAP}$ is slightly overstructured 
takes account of both the experimental uncertainties and the quantum nuclear effects.


\section{Discussion and conclusions}
\label{sec:discussion}

The machine-learning methods we have described offer a new approach to the
modelling of molecular materials. To show the possibilities, we have discussed only a single
material (water) and we have focused on the systematic improvement of the 1- and 2-body parts
of the energy. Furthermore, the use of machine learning to represent the difference between
DFT and accurate quantum chemistry (and thus incurring the cost of DFT when running a simulation) 
is only one of several possible strategies. There is
ample scope for removing these limitations in the future. Note, however, what has already
been achieved even with these limitations. The description of water energetics for aggregation
states ranging from clusters to extended solid and liquid states is challenging because
of the delicate balance between different components of the energy. We have shown how
to use machine learning to construct a quantitative correction for 1-, 2-body
components of the energy, starting from a chosen DFT approximation (in this case, BLYP). We have
achieved a substantial improvement in the description of cluster and ice-phase
energetics and liquid-state properties, with a completely negligible increase in computational cost, and we have also revealed a remaining inaccuracy
in the beyond-2-body energetics, which we attribute to exaggerated cooperativity of
H-bonding.

The most obvious generalisation now needed is to use machine learning to systematically
improve the beyond-2-body energetics. In fact, GAP-based machine learning has already been
used with great success to represent many-body energies in other types of systems~\cite{bartok10},
and there is every reason to expect that this can be done for molecular systems. As a first
step, we plan to use accurate beyond-2-body energies generated for large samples
of configurations of small and moderate sized clusters, as we have already done in a recent paper~\cite{gillan2012}.
Benchmark information from quantum Monte Carlo for configuration samples of larger clusters
and for the liquid may also be helpful.

Our strategy of using GAP to represent corrections to be added to DFT
is powerful in terms of generality, but large
gains in computational efficiency could be achieved with other strategies. The use of
GAP to represent corrections to empirical force fields, particularly those that already
account for molecular flexibility and
for many-body effects (see e.g. Ref.~\cite{fanourgakis2008,hasegawa2011,babin2012}), holds 
much promise for the future. However, we note that although such models already exist for water,
as a result of many years of development effort, the same is true of very few other molecular systems.

We have focused here on water, because of its outstanding importance for science and society,
but the present methods should be readily applicable to many other molecular systems,
even without the technical improvements we have mentioned. The cluster, solid and
liquid states of simple molecules such as ammonia and hydrogen fluoride are obvious
targets, but many kinds of mixture, including the environmentally and
industrially important gas hydrates, are also accessible.

\section {Materials and Methods}
All calculations on molecules and isolated molecular assemblies were carried out using the Molpro package\cite{molpro}. The water dimer configurations were obtain from a long equilibrium simulation of liquid water using the AMOEBA force field\cite{ren2003}. 

We used the Castep code~\cite{castep} for the DFT calculations on ice structures, and a modified version of the VASP code \cite{vasp} for the molecular dynamics simulations of liquid heavy water (D$_2$O). Here, the number of molecules per unit volume is the same as for H$_2$O at $0.997$~g~cm$^{-3}$. Further technical details about basis-set completeness, time-step, temperature control etc are given in the Supplementary Information (SI), but we note that thermal averages were computed on a run of 45~ps duration of which  20~ps was discarded for equilibration.

Software and data are available at http://www.libatoms.org.

\section{Acknowledgements}
ABP was supported by a Junior Research Fellowship at Magdalene College, Cambridge. 
GC acknowledges supported from the Office of Naval Research under grant number N000141010826, and EU-FP7-NMP grant 229205 ADGLASS. The authors are grateful to D~O'Neill for his important contribution to an earlier incarnation of this project, and FRM gratefully acknowledges funding from the Engineering and Physical Sciences Research Council (EP/F000219/1). The authors thank AK~Soper and CJ~Benmore for useful discussions.

\begin{figure}
\begin{center}
\includegraphics[width=8cm]{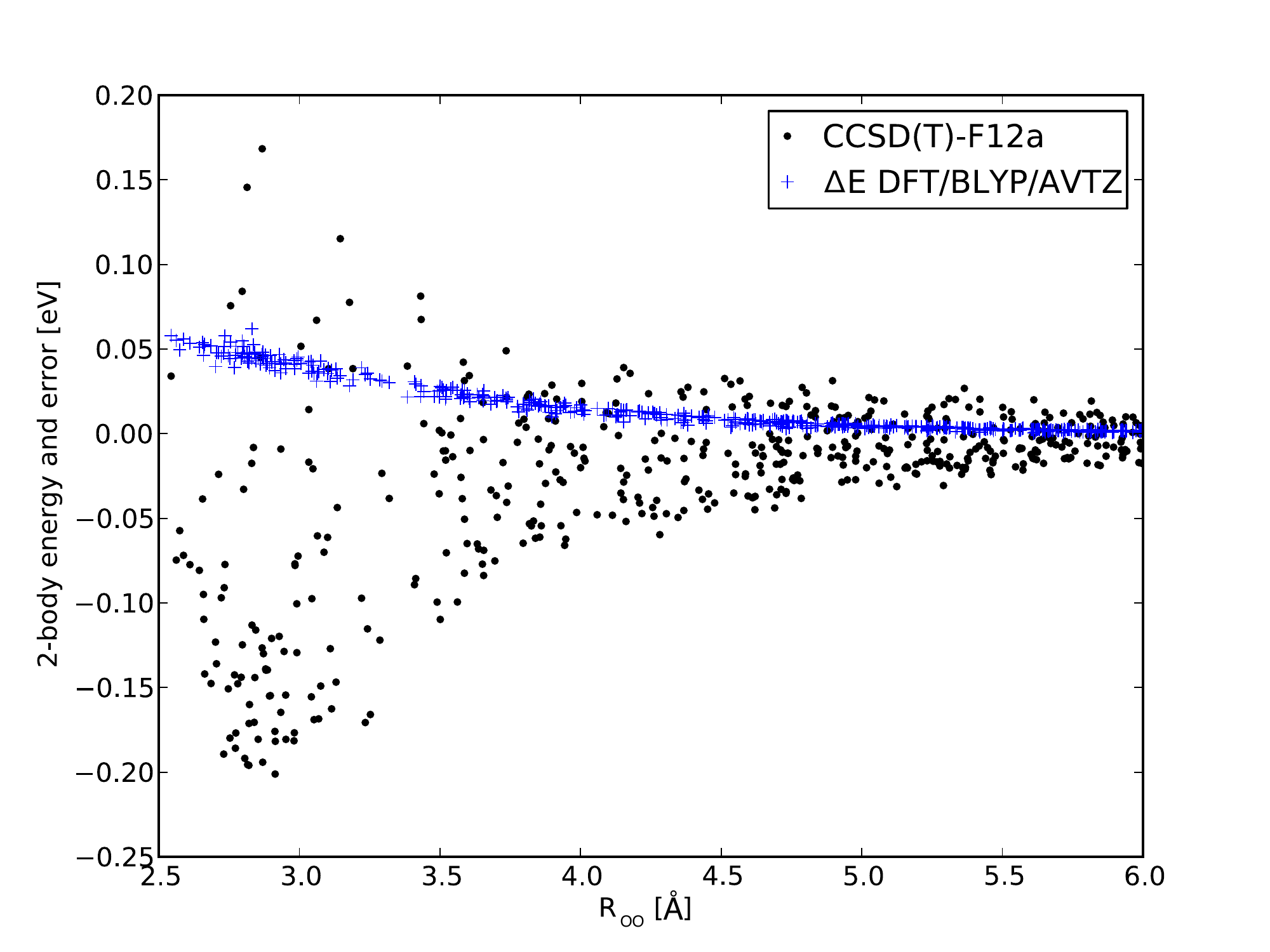}
\includegraphics[width=8cm]{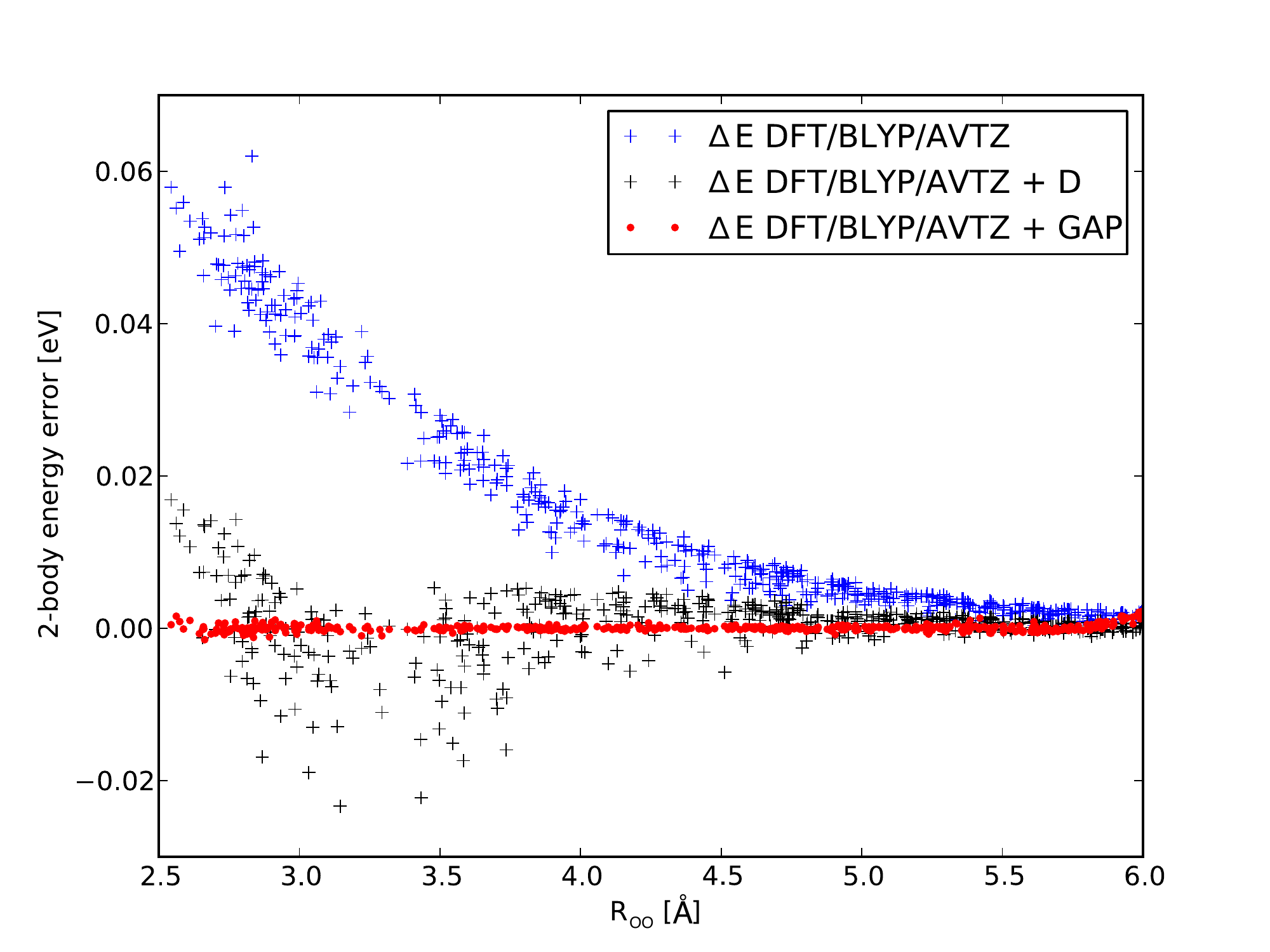}
\end{center}
\caption{Top: Benchmark 2-body interaction (black) and errors of DFT with BLYP functional (blue) plotted against
oxygen-oxygen distance $R_{\rm OO}$. Benchmarks are computed with CCSD(T) close to the basis-set limit.
Bottom: 2-body errors of BLYP (blue, same as in top panel) and errors of BLYP+GAP (red). The RMS
deviation of BLYP+GAP from benchmarks is $0.45$~meV. Also shown are errors of BLYP
plus Grimme D3 dispersion correction~\cite{grimme2010}. The sample of 500 dimer configurations shown here were drawn from a molecular dynamics simulation of liquid water and not used in the construction of the GAP models.}
\end{figure}

\begin{figure}
\includegraphics[width=8cm]{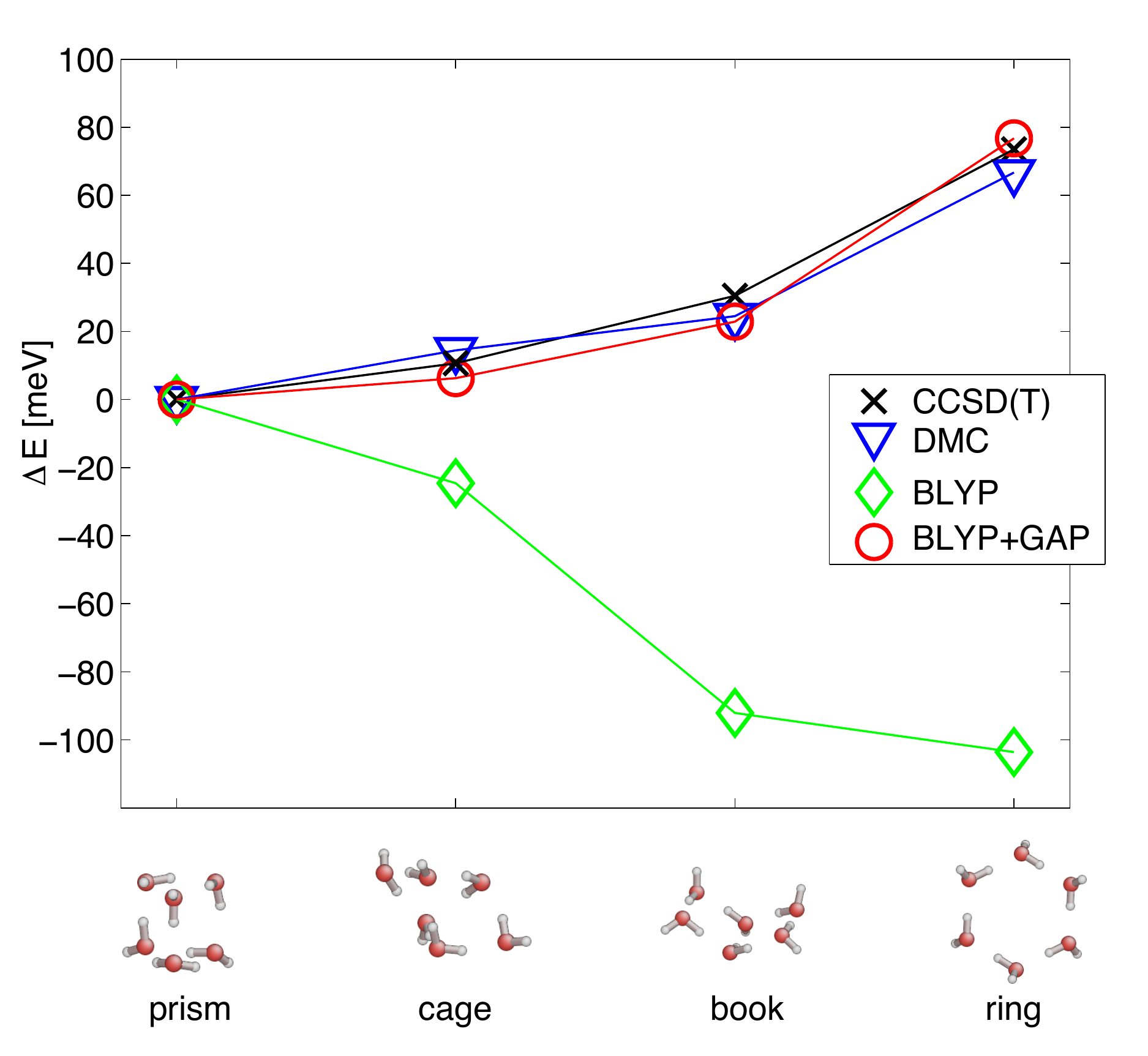}
\caption{Relative energies (meV units) of four isomers of the water hexamer (atomic coordinates from
Ref.~\cite{santra2008}) computed with CCSD(T) close to the basis-set limit~\cite{bates2009},
diffusion Monte Carlo~\cite{santra2008}, BLYP and BLYP+GAP. Geometries are depicted 
below the Figure.}
\label{fig:hex}
\end{figure}

\begin{figure}
\begin{center}
\includegraphics[width=8cm]{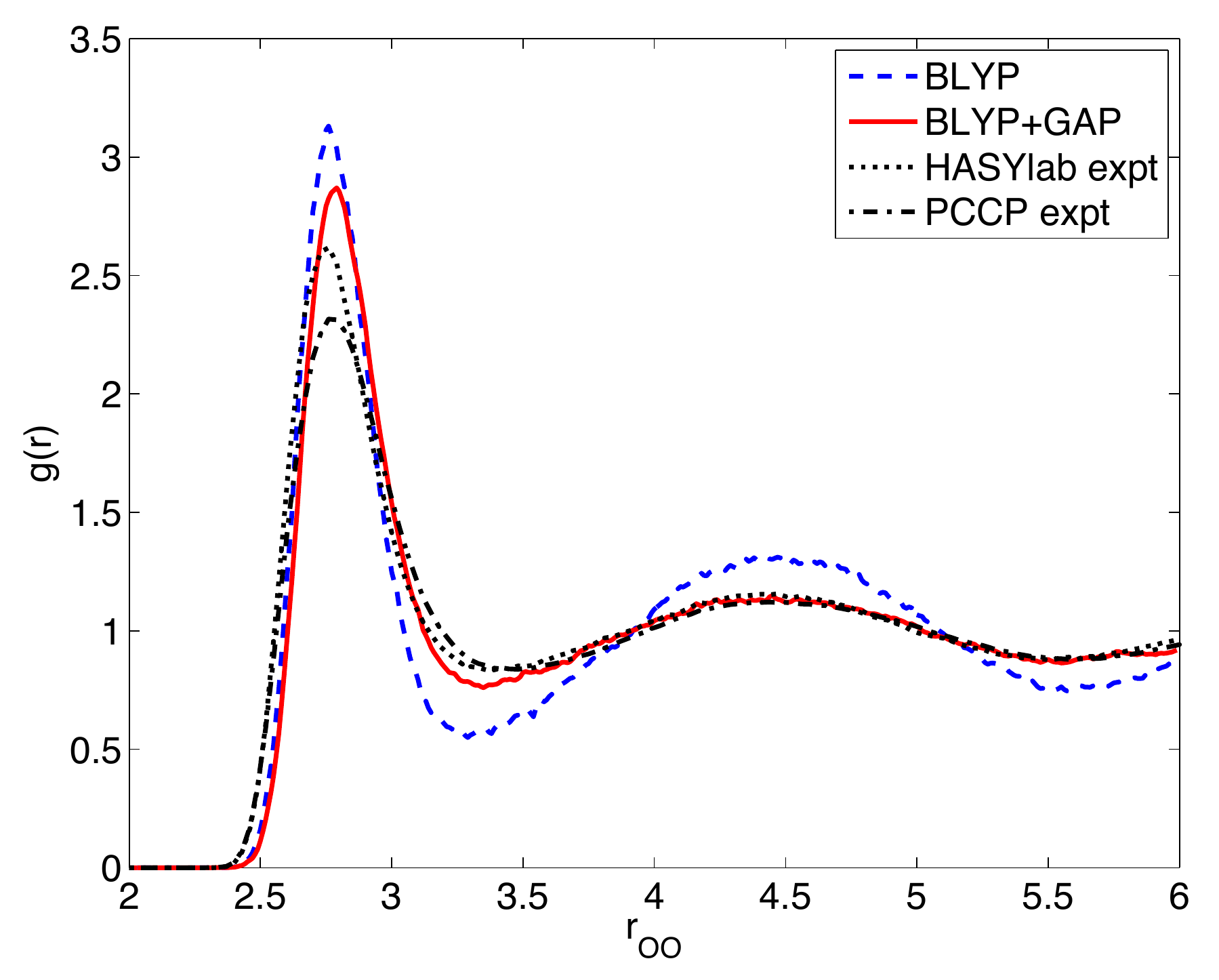}
\end{center}
\caption{Oxygen-oxygen radial distribution function of liquid water at $308$~K at experimental density
using BLYP (blue dashed) and BLYP+GAP (red solid) compared with two sets of experimental data
(black dotted and dash-dotted). Experimental data are from joint refinement of neutron data
and two sets of x-ray data, identified as HASYlab and PCCP (see Ref.~\cite{soper2007} for details).}
\label{fig:goo}
\end{figure}

\begin{table}
\caption{Relative energies of the 10 stationary points of the water dimer computed with CCSD(T) close to
the basis-set limit~\cite{tschumper2002}, diffusion Monte Carlo~\cite{gillan2012}, DFT with the BLYP functional
and BLYP+GAP. Numbering of the stationary states is standard (see e.g. Ref.~\cite{tschumper2002}). 
Units: meV.}
\label{tab:dimstat}
\begin{tabular}{crrrr}
State & BLYP & BLYP+GAP & CCSD(T) & DMC  \\\hline
1  &    0 &    0&   0  &    0  \\
2  &   23 &   23&  21  &   24  \\
3  &   32 &   27&  25  &   27  \\
4  &   49 &   32&  30  &   34  \\
5  &   65 &   44&  41  &   39  \\
6  &   73 &   45&  44  &   41  \\
7  &   96 &   79&  79  &   78  \\
8  &  155 &  153& 154  &  156  \\
9  &   95 &   82&  77  &   79  \\
10\phantom. &  125 &  116& 117  &  122  \\
\end{tabular}
\end{table}

\begin{table}
\caption{Binding energies and volumes of ice polymorphs computing using DFT with the BLYP
functional and BLYP+GAP compared with experimental values~\cite{soper2007}.
Zero-point vibrational contributions have been removed from the experimental
energies~\cite{whalley1984}, but not from the volumes. 
}
\label{tab:ice}
\begin{tabular}{lrrrrrr}
&Ih&II&VIII&IX&XI&XV\\
\hline
{\em Binding energy} [meV]\\
BLYP&$-$540&$-$458&$-$318&$-$475&$-$544&$-$403\\
BLYP+GAP&$-$667&$-$672&$-$637&$-$670&$-$671&$-$657\\
EXPT&$-$610&$-$609&$-$579&$-$606&&\\
\hline
{\em Binding energy relative to Ih} [meV]\\
BLYP&0&83&223&66&$-$3&138\\
BLYP+GAP&0&$-$5&30&$-$3&$-$4&$-$11\\
EXPT&0& 1&31&4&&\\
\hline
\hline
{\em Volume }  $[\text{\AA}^3]$\\
BLYP&31.7&26.0&21.2&28.0&32.1&24.6\\
BLYP+GAP&30.6&23.8&18.6&24.0&30.6&21.1\\
EXPT&32.0&25.3&20.1&25.6&32.0&22.9\\
\hline
{\em Volume relative to Ih} $[\text{\AA}^3]$\\
BLYP&0&$-$5.7&$-$10.5&$-$3.8&0.36&$-$7.9\\
BLYP+GAP&0 &$-$6.7&$-$11.9&$-$6.5&0&$-$9.5\\
EXPT&0&$-$6.7&$-$11.9&$-$6.4&0&$-$9.2\\
\end{tabular}
\end{table}

\end{document}